\documentclass[aps,12pt,pra,showpacs,preprint,a4paper]{revtex4}
\usepackage{amsmath}
\usepackage{graphicx}
\usepackage{subfig}
\usepackage{array}
\usepackage{amsfonts}
\usepackage{epsf}
\usepackage{epsfig}
\usepackage{amssymb}

\begin{document}
\title{Bound state solutions of square root power law potential--Wavefunction ansatz method on $D$-dimensional Schr\"{o}dinger equation}
\author{\small Tapas Das}
\email[E-mail: ]{tapasd20@gmail.com}\affiliation{Kodalia Prasanna Banga High
 School (H.S), South 24 Parganas, 700146, India}
\begin{abstract}
$D$-dimensional Schr\"{o}dinger equation is addressed for square root power law potential. Bound state unnormalized eigenfunctions and the energy eigenvalues are obtained using wave function ansatz method. Some special cases are studied at the end to ensure the correctness of present work.
\end{abstract}
\maketitle
\newpage
\section{I\lowercase{ntroduction}}
Over the past few years, theoretical physicists have shown a great deal of interest in solving higher dimensional
 Schr\"{o}dinger equation for various potentials[1-12]. Apart from regular well known potentials, also some complicated potential like ring shaped potential coupled with Coulomb potential, exponential potential, hyperbolic potential or their different combinations are also addressed by many authors [13-19].\\
Unfortunately, in this part of literature so far the solutions of higher dimensional Schr\"{o}dinger equation with anharmonic potentials like combination of singular fractional power potential with \textit{Singular integer power potential} $V(r)=\sum_{i=0}^p \frac{a_i}{r^i}$, \textit{Sextic potential} $V(r)=ar^6+br^4+cr^2$, \textit{Non-polynomial potential} $V(r)=r^2+\frac{\lambda r^2}{1+gr^2}$ and others have not yet been paid a considerable amount of interest.Singular fractional power law potential has been recently found application in quark-antiquark interaction [20] and in models of exhibiting shape resonance behavior [21]. There are few number of studies basically on lower dimensional Schr\"{o}dinger equation via various approximation method[22-25]. Motivated by the lack of these type of studies in higher dimensional Schr\"{o}dinger equation, an attempt has been maid in this paper to solve the following square root power potential  
\begin{eqnarray}
V(r)=\sum_{i=0}^4\frac{a_i}{r^{\frac{i}{2}}}\,,
\end{eqnarray}
where $a_i(i=0,1,2,3,4)$ represent the potential parameters. The potential under consideration is not purely fractional power potential rather than it is a combination of singular fractional power potential with singular integer power potential. It is clear due to the mathematical difficulty only the running index is taken up to $i=4$. At present there are number of suitable techniques available [26-29] to solve different types of differential equation that generally appears in physical systems like Schr\"{o}dinger equation in quantum mechanics. However, in this paper traditional wave function ansatz concept [30-31] with a little series solution technique[32] has been employed to solve the potential for multidimensional Schr\"{o}dinger equation. \\
The paper is organized as follows: Next section is for the bound state spectrum of the potential. In Section III special cases are studied. Section IV is for the conclusion of the work and finally important references are given in section V.
\newpage 
\section{B\lowercase{ound} s\lowercase{tate} s\lowercase{pectrum} \lowercase{of the} P\lowercase{otential}}
The $D$-dimensional time-independent Schr\"{o}dinger equation for a particle of mass $M$ with arbitrary angular quantum number 
$\ell$ is given by [33] (in units of $\hbar=1$)
\begin{eqnarray}
\Big[\nabla_D^{2}+2M\Big(E_{n\ell}-V(r)\Big)\Big]\psi_{n\ell m}(r,\Omega_D)=0\,,
\end{eqnarray}
where $E_{n\ell}$ and  $V(r)$ denote the energy eigenvalues and potential. $\Omega_D$ within the argument of $n$-th state eigenfunctions $\psi$ denotes angular variables
$\theta_1,\theta_2,\theta_3,.....,\theta_{D-2},\varphi$. The Laplacian operator in hyperspherical coordinates is written as 
\begin{eqnarray}
\nabla_D^{2}=\frac{1}{r^{D-1}}\frac{\partial}{\partial r}(r^{D-1}\frac{\partial}{\partial r})-\frac{\Lambda_{D-1}^{2}}{r^2}\,,
\end{eqnarray}
where 
\begin{eqnarray}
\Lambda_{D-1}^{2}=-\Bigg[\sum_{k=1}^{D-2}\frac{1}{sin^2\theta_{k+1}sin^2\theta_{k+2}.....sin^2\theta_{D-2} sin^2\varphi}\times\left(\frac{1}{sin^{k-1}\theta_k}\frac{\partial}{\partial \theta_k}sin^{k-1}\theta_k\frac{\partial}{\partial \theta_k}\right)\nonumber\\+\frac{1}{sin^{D-2}\varphi}\frac{\partial}{\partial\varphi}sin^{D-2}\varphi\frac{\partial}{\partial\varphi}\Bigg]\,.
\end{eqnarray}
$\Lambda_{D-1}^{2}$ is known as hyperangular momentum operator.\\ 
Expressing the solution as $\psi_{n\ell m}(r,\Omega_D)=R_{n\ell}(r)Y_\ell^{m}(\Omega_D)$ and applying the separation variable method in Eq.(2) we can have two separated equations 
\begin{eqnarray}
\Lambda_{D-1}^{2}Y_\ell^{m}(\Omega_D)&=\ell(\ell+D-2)Y_\ell^{m}(\Omega_D)\,,
\end{eqnarray} 
and
\begin{eqnarray}
\Bigg[\frac{d^2}{dr^2}+\frac{D-1}{r}\frac{d}{dr}-\frac{\ell(\ell+D-2)}{r^2}+\frac{2M}{\hbar^2}\Big\{E_{n\ell}-V(r)\Big\}\Bigg]R_{n\ell}(r)=0\,,
\end{eqnarray}
where $Y_\ell^{m}(\Omega_D)$ in the Eq.(5) are the hyperspherical harmonics. $\ell$ is called the orbital angular momentum quantum number, where $\ell=0,1,2,3....$.\\Eq.(6) is called the hyperradial or in short ``radial'' equation containing the radial part of the solution $R_{n\ell}(r)$. The separation constant $\ell(\ell+D-2)$ is only applicable for $D>1$. \\
Now inserting the Eq.(1) into Eq.(6) and changing the variable as $r^{\frac{1}{2}}=x$ we have
\begin{eqnarray}
\Bigg[\frac{d^2}{dx^2}+\frac{2D-3}{x}\frac{d}{dx}-\frac{N_{\ell a_4}^D}{x^2}+8M\Big\{(E_{n\ell}-a_0)x^2-a_1x-a_2-\frac{a_3}{x}\Big\}\Bigg]R(x)=0\,,
\end{eqnarray}
where $N_{\ell a_4}^D=4\ell(\ell+D-2)+8Ma_4$ and $R_{n\ell}(r)\leftrightarrow R(x)$.\\
In order to solve Eq.(7) we first find the asymptotic behavior of $R(x)$. As $x\rightarrow\infty$ we have
\begin{eqnarray}
\frac{d^2R(x)}{dx^2}+8M\Big\{(E_{n\ell}-a_0)x^2-a_1x\Big\}R(x)=0\,.
\end{eqnarray}
The solution must be bounded at infinity i.e $R(x\rightarrow\infty)\rightarrow 0$. So let us try the solution of Eq.(8) as
\begin{eqnarray}
R(x)=[const] e^{-(Ax^2+Bx)}\,,
\end{eqnarray}
where $A,B$ are constants. Taking this form of solution in Eq.(8) and neglecting the term $B^2-2A$, as it is fare for large limit of $x$, we have
by comparing the coefficient of $x^2$ and $x$ 
\begin{subequations}
\begin{align}
A&=\sqrt{-2M(E_{n\ell}-a_0)}\,,\\
B&=\frac{2Ma_1}{\sqrt{-2M(E_{n\ell}-a_0)}}\,.
\end{align}
\end{subequations}
The solution must be bounded in origin also. The well-behaved complete solution of $R(x)$ now can be taken as
\begin{eqnarray}
R(x)=[const]e^{-(Ax^2+Bx)}f(x)\,, 
\end{eqnarray}
were $f(x\rightarrow 0)\rightarrow 0$.\\
Using Eq.(11), Eq.(7) leads to the following homogeneous linear second order differential equation
\begin{eqnarray}
\frac{d^2f(x)}{dx^2}+h(x)\frac{df(x)}{dx}+g(x)f(x)=0\,,
\end{eqnarray}
where
\begin{subequations}
\begin{align}
h(x)&=\frac{2D-3}{x}-(4Ax+2B)\,, \\
g(x)&=(B^2-2A-8Ma_2)-\frac{8Ma_3+(2D-3)(2Ax+B)}{x}-\frac{N_{\ell a_4}^D}{x^2}\,.
\end{align}
\end{subequations}
The complete solution of this differential equation can be found if one integral in its complementary function be known.
Let us try an inspection such that $f(x)=x^k$ is a solution of this equation. So we have
\begin{eqnarray}
k(k-1)+H(x)k+G(x)=0\,,
\end{eqnarray}
where $H(x)=xh(x)$ and $G(x)=x^2g(x)$.\\
Now $h(x)$ and $g(x)$ are not analytic near the origin, but $H(x)$ and $G(x)$ are both analytic at $x=0$.This provides
$k(k-1)+H(0)k+G(0)=0$ and hence
\begin{subequations}
\begin{align}
k_+&=-(D-2)+\sqrt{(D+2\ell-2)^2+8Ma_4}\,,\\
k_-&=-(D-2)-\sqrt{(D+2\ell-2)^2+8Ma_4}\,.
\end{align}
\end{subequations}
The acceptable solution must be $k_+(>0)$ given by Eq.(15a) because it preserves the rule $f(x\rightarrow 0)\rightarrow 0$.\\
Now we are in a situation to write the form of complete solution for $f(x)$ as a series like
\begin{eqnarray}
f(x)=x^k\sum_{p=0}^{\infty}c_px^p\,.
\end{eqnarray}
Taking this into Eq.(12) we have the following identity
\begin{eqnarray}
\sum_{p=0}^{\infty}\alpha_p c_px^{k+p+2}-\sum_{p=0}^{\infty}\beta_p c_px^{k+p+1}+\sum_{p=0}^{\infty}\gamma_p c_px^{k+p}=0\,,
\end{eqnarray}
where
\begin{subequations}
\begin{align}
\alpha_p&=B^2-8Ma_2-4A(D+k+p-1)\,, \\
\beta_p&=8Ma_3+B(2D+2k+2p-3)\,,\\
\gamma_p&=(k+p)(k+p+2D-4)-N_{\ell a_4}^D\,.
\end{align}
\end{subequations}
Equating the coefficient of lowest power of $x$, which corresponds to $p=0$ in the left hand side of the identity, we have the indicial equation
\begin{eqnarray}
c_0[k(k-1)+k(2D-3)-N_{\ell a_4}^D]=0\,.
\end{eqnarray} 
We must choose $c_0\neq 0$ as $k=k_+$ makes the bracket term of Eq.(19) a vanishing. \\
Similarly equating the coefficient of next higher power of $x$, i.e $x^{k+1}$ from the identity we get
\begin{eqnarray}
c_1=\frac{8Ma_3+B(2k+2D-3)}{(k+1)(k+2D-3)-N_{\ell a_4}^D}c_0\,. 
\end{eqnarray} 
After deciding the coefficients $c_0$ and $c_1$, finally equating the coefficient of $x^{p+k+2}$ from both sides of the given identity i.e Eq.(17) we have
\begin{eqnarray}
\alpha_pc_p-\beta_{p+1}c_{p+1}+\gamma_{p+2}c_{p+2}=0\,.
\end{eqnarray}  
Now for physical cases the series must be a polynomial or in short it must terminate at $p=n$. That means we must have
$c_n\neq 0$ and $c_{n+1}, c_{n+2}, c_{n+3},...=0 $\\
From the above argument it is easy to find $\alpha_n=0$ which finally provides the energy eigenvalue equation in association with Eq.(10b) as
\begin{eqnarray}
4A^3(D+k+n-1)+8A^2Ma_2-4M^2a_1^{2}=0\,.
\end{eqnarray}
The corresponding radial eigenfunctions are
\begin{eqnarray}
R_{n\ell}(r)=N_{n\ell}e^{-\Big[r\sqrt{-2M(E_{n\ell}-a_0)}+r^{1/2}\frac{2Ma_1}{\sqrt{-2M(E_{n\ell}-a_0)}}\Big]}r^{\frac{k_+}{2}}\sum_{j=0}^{n}c_j r^{j/2}\,,
\end{eqnarray}
where we have used Eq.(10a, 10b),Eq.(11) and Eq.(16) with the exchange of variable $x=r^{1/2}$. $N_{n\ell}$ is the normalization constant which can be evaluated from the condition $\int_{0}^{\infty}[R_{n\ell}(r)]^2r^{D-1}dr=1$.
\section{D\lowercase{iscussion of} s\lowercase{pecial} c\lowercase{ases}}
\begin{enumerate}
\item{\bf{Singular One Fractional Power Potential: $V(r)=\frac{a_1}{r^{1/2}}+\frac{a_3}{r^{3/2}}$}}\\
The potential is obtained from Eq.(1) by setting $a_0=a_2=a_4=0$ \\ 
Under this condition from Eq.(15a) we have $k_+=2\ell$, and from\\ Eq.(10a) $A=\sqrt{-2ME_{n\ell}}$. \\
Now Eq.(22) immediately gives the energy eigenvalues
\begin{eqnarray}
E_{n\ell}=-\frac{1}{2M}\Bigg(\frac{M^2a_1^2}{2k^{'}+n+1}\Bigg)^{\frac{2}{3}}\,,
\end{eqnarray}
where $k^{'}=(\frac{D}{2}+\ell-1)$. \\
The corresponding radial eigenfunctions are given by
\begin{eqnarray}
R_{n\ell}(r)=N_{n\ell}e^{-\Big[r\sqrt{-2ME_{n\ell}}+r^{1/2}\frac{2Ma_1}{\sqrt{-2ME_{n\ell}}}\Big]}r^{\ell}\sum_{j=0}^{n}c_j r^{j/2}\,.
\end{eqnarray}
The above results are in excellent agreement with the ref.[33].
\item{\bf{Mie-type Potential: $V(r)=a_0+\frac{a_2}{r}+\frac{a_4}{r^2}$}}\\
The potential is obtained from Eq.(1) by setting $a_1=a_3=0$.\\
Eq.(10b) gives $B=0$ hence Eq.(18b) provides $\beta_p$ (or $\beta$)$=0$.\\ 
Under this condition Eq.(20) gives $c_1=0$ and the recurrence relation comes out from Eq.(21) as
\begin{eqnarray}
\alpha_pc_p+\gamma_{p+2}c_{p+2}=0\,.
\end{eqnarray}
The above relation immediately gives
\begin{eqnarray}
c_p\Rightarrow \begin{cases} =0, & \text{if }  p\text{ is odd,}\\
\neq 0, & \text{if }  p \text{ is even.}\end{cases}
\end{eqnarray}
Taking $p=2n_r$ where $n_r=0,1,2,3.....$ we again argue that the series must terminate at $p=2n_r$ (i.e $c_{2n_r}\neq 0$) to achieve a polynomial solution to explain the physical cases. So under this situation in Eq.(26) we have to set $c_{p+2}=0$ for $p>2n_r$.\\
This gives $\alpha_{2n_r}=0$ which is nothing but the same as Eq.(22) except $p\equiv n=2n_r$ and potential parameter $a_1=0$.\\
Hence the energy eigenvalues are
\begin{eqnarray}
E_{n_r\ell}=a_0-\frac{M}{2}\Bigg[\frac{a_2}{n_r+\frac{1}{2}\Big\{1+\sqrt{(D+2\ell-2)^2+8Ma_4}\Big\}}\Bigg]^2\,.
\end{eqnarray}
This is what obtained in ref.[12] and ref.[31].\\
The radial eigenfunctions for this case emerges from Eq.(23) as
\begin{eqnarray}
R_{n_r\ell}(r)=N_{n_r\ell}e^{-\Big[r\sqrt{-2M(E_{n_r\ell}-a_0)}\Big]}r^{\frac{k_+}{2}}\sum_{\stackrel{j=0} {j\neq odd}}^{2n_r}c_j r^{j/2}\,.
\end{eqnarray}
\item{\bf{Coulomb Potential:} $V(r)=\frac{a_2}{r}$}\\
This can be obtained from Eq.(1) by setting $a_0=a_1=a_3=a_4=0$. Thus the results of Mie-type potential as well as the explanations of that section are equally applicable and can be used here with an extra condition $a_0=a_4=0$.
So the energy eigenvalues are
\begin{eqnarray}
E_{n_r\ell}=-\frac{M}{2}\Bigg(\frac{a_2}{\frac{D}{2}+\ell+n_r-\frac{1}{2}}\Bigg)^2\,.
\end{eqnarray} 
This result is similar to the ref.[10]. \\
Eq.(15a) provides $k_+=2\ell$, hence we have the radial eigenfunctions from Eq.(29) as
\begin{eqnarray}
R_{n_r\ell}(r)=N_{n_r\ell}e^{-r\Big[\sqrt{-2ME_{n_r\ell}}\Big]} r^{\ell}\sum_{\stackrel{j=0} {j\neq odd}}^{2n_r}c_j r^{j/2}\,.
\end{eqnarray}
\end{enumerate}
\section{C\lowercase{onclusions}}
In this paper $D$-dimensional Schr\"{o}dinger equation is addressed for square root power law potential using wave function ansatz method.  It has been shown that traditional mathematical method like wave function ansatz and series solution always provide a wide space to deal with complicated potential problems. The result that has been obtained in this work is satisfactory in the sense that some special cases like singular one fractional power potential, Mie-type potential, Coulomb potential are derived easily and they are in excellent agreement with the previous works. \\
The exact form of the energy eigenvalues given by Eq.(22) is complicated. The only way to find the energy eigenvalues is the numerical computation of the equation for a particular state. However the eigenfunctions are in well known form. May be after normalization they can be expressed in terms of Laguerre polynomials or confluent hypergeometric functions.\\
We will look forward to the further study of this potential in future at least for few more terms ($i>4$). Any other method like asymptotic iteration method (AIM), Nikiforov-Uvarov method (NU) will be helpful to this. \\
Furthermore, following the same way of this paper, one can study the other class of fractional singular power potentials like 
$\sum\frac{a_i}{r^{i/3}}$ and $\sum\frac{a_i}{r^{i/4}}$. 
\section{R\lowercase{eferences}}

\end{document}